\documentclass[11pt]{article} 

\usepackage[utf8]{inputenc} 

\usepackage{geometry} 
\usepackage{graphicx} 

\usepackage{booktabs} 
\usepackage{array} 
\usepackage{paralist} 
\usepackage{verbatim} 
\usepackage{subfig} 
\usepackage{jheppub}
\usepackage[utf8]{inputenc}
\usepackage{fancyhdr} 
\usepackage{titlesec}

\titleformat{\section}
       {\fontsize{12}{17}\bfseries}{\thesection}{1em}{}

\titleformat{\subsection}
       {\fontsize{12}{17}\bfseries}{\thesubsection}{1em}{}

\usepackage[nottoc,notlof,notlot]{tocbibind} 
\usepackage[titles,subfigure]{tocloft} 

\usepackage{amssymb}

\AtBeginDocument{}

\title{Emergent Dark Gravity from (Non)Holographic Screens}
\author{Alexander Peach}
\affiliation{Durham University Department of Physics,\\
South Rd, Durham, UK}
\emailAdd{a.m.peach@durham.ac.uk}
\abstract{In this work, a clear connection is made between E. Verlinde's recent theory of emergent gravity in de Sitter space \cite{verl2016}, and the earlier work describing emergent gravity using holographic screens \cite{verl2010}. A modified (non)holographic screen scenario is presented, wherein the screen fails to encode an emergent mass in the bulk ``unemerged" part of space for sufficiently large length-scales, where the volume-law of the non-holographic bulk degrees of freedom overtakes the area-law scaling of the entropy of the screen. Within this framework, we can describe both an emergent dark gravitational force, which scales like $\frac{1}{r}$, and also a version of the baryonic Tully-Fisher relation. We therefore recast these results within an emergent gravity framework in which there is an explicit violation of holography for sufficiently large length-scales.}
\begin{document}
\maketitle

\section{Introduction}

It is widely assumed that our understanding of gravity according to Einstein, together with effective local quantum field theories give an accurate description of IR physics. On the other-hand, many recent insights within the fields of holography and entropic gravity continue to corroborate the notion that spacetime and the laws of gravity should emerge from a UV-complete theory of quantum gravity\cite{J95, M01, VR2010, susskind13, pad15, susskind17, felix17}\nocite{smc} \nocite{vl18}
\nocite{Heem}. It has become clear, particularly in the context of AdS/CFT, that the emergence of the known laws of gravity is intimately connected with an entanglement area-law \cite{ryu06, hubeny, myers13, swingle14, C16, C17}. Conversely, deviations from the area-law entanglement could lead to violations of gravitational physics. 

That there could be novel physical ramifications of high-energy physics upon the laws that govern arbitrarily-large scales might seem very surprising, but just such a connection was recently proposed by E. Verlinde in the recent work \cite{verl2016}. This recent work offers a novel hypothesis for the relation between the de Sitter entropy associated with the cosmological horizon, in terms of the dark energy, and its response due to the addition of matter. It is argued that the entropy carried by the dark energy obeys a volume-law, and that adding matter to a region of space generates competition between the volume-law and the area-law entropy associated with the matter. It is shown that there is a length-scale, which is related to the mass and the curvature scale, where the volume-law for the dark energy overwhelms the area-law scaling of the matter entropy. This leads to a violation of the holographic principle above this length-scale. It is this key role played by the violation of holography, at sufficiently large length-scales that is highlighted in the present work. 

In \cite{verl2016}, it was shown that the interplay between the de Sitter entropy, carried by the dark energy, and its local reduction due to the addition of a mass $M$, pertains to the following criteria,

\begin{equation}
\label{criteria}
\frac{M}{A(R)} \lessgtr \frac{a_0}{8 \pi G},
\end{equation}
where $a_0 = \frac{ c^2 }{ L }$ is the MOND acceleration scale, $L$ is the Hubble length and $A(R)$ is the area of a sphere with radius $R$. This entropic condition coincides exactly with the empirical criteria for the scale upon which the phenomena attributed to dark matter become manifest in galactic rotation curves. This empirical observation is then given physical motivation by Verlinde's new hypothesis. Furthermore, it is shown that one can reproduce many key empirical findings in cosmology, such as Milgrom's fitting formula of MOND and the baryonic Tully-Fisher relation within this formalism \cite{milgrom, btf}. There is also strong agreement between this framework and weak-lensing observations \cite{brouwer16}.

The paper \cite{verl2016} has been criticised on the basis of apparent inconsistencies in terms of associating an effective elastic description to the response governed by the removal (by the addition of matter) of a certain volume of the dark energy medium. Several of these features have since been addressed in the recent paper \cite{sabine}, where a covariant proposal for the effective elastic response of the dark energy was considered, and the roles of the identifications between gravitational and elastic variables were clarified. Since the elastic description is intended to give an effective description, its negation would not necessarily rule out the underlying postulates of the emergent gravity framework. In the present work we make contact with the key results of \cite{verl2016}, without using an effective elastic description.

The main proposal of the present paper is to use a straightforward modification of the setup described in \cite{verl2010}, where Newtonian and Einstein gravity were derived from arguments involving so-called holographic screens. The modification entails a violation of holography above a critical length-scale, which thereby emulates a key feature of Verlinde's new theory. In the present paper, it is demonstrated that this modification is sufficient to describe the emergence of a dark gravity force, as well as a version of the baryonic Tully-Fisher relation. This work does not claim to go further than Verlinde's proposals, but instead the present goal is to provide a framework for the underlying ideas that is mathematically simpler, and which clarifies the key role played by the breakdown of holography in the emergence of dark gravity.

The rest of this paper is organised as follows. Section \ref{review} contains a review of the recent proposal \cite{verl2016} which identifies the core components that will be utilised in the derivation of emergent dark gravity, presented in section \ref{main}. In section \ref{main}, the main result is presented, where a modified thermodynamic setup based on arguments involving a holographic screen is constructed. It is demonstrated that the emergent entropic force associated to this system receives an additional contribution from the information in the bulk which is no-longer encoded holographically. This additional force has a $r^{-1}$ scaling-law required to describe flattening galactic rotation curves. In this setup it is also shown that the bulk and boundary energies obey a relation which is analogous to the baryonic Tully-Fisher relation, and that the bulk energy manifests as an apparent mass. By inputting the critical scale $r_c(M,L)$ identified in Verlinde's new theory to the setup, we find that the entropic force associated to the bulk then takes exactly the form of the force required to describe flat galactic rotation curves. It is also found that this value for the critical scale reproduces the baryonic Tully-Fisher relation up to a numerical factor. The key assumptions that are used in the present work are then reviewed. Finally, in section \ref{relation} the present work is compared to the work of several recent papers with related goals.

\section{Emergent Gravity in de Sitter Space}
\label{review}

In \cite{verl2016} a radical new explanation for the phenomena attributed to dark matter was offered in terms of emergent gravity in de Sitter space. In this section, we briefly review the key features of this work.

\begin{figure}
\centering
\includegraphics[width=0.8\textwidth]
{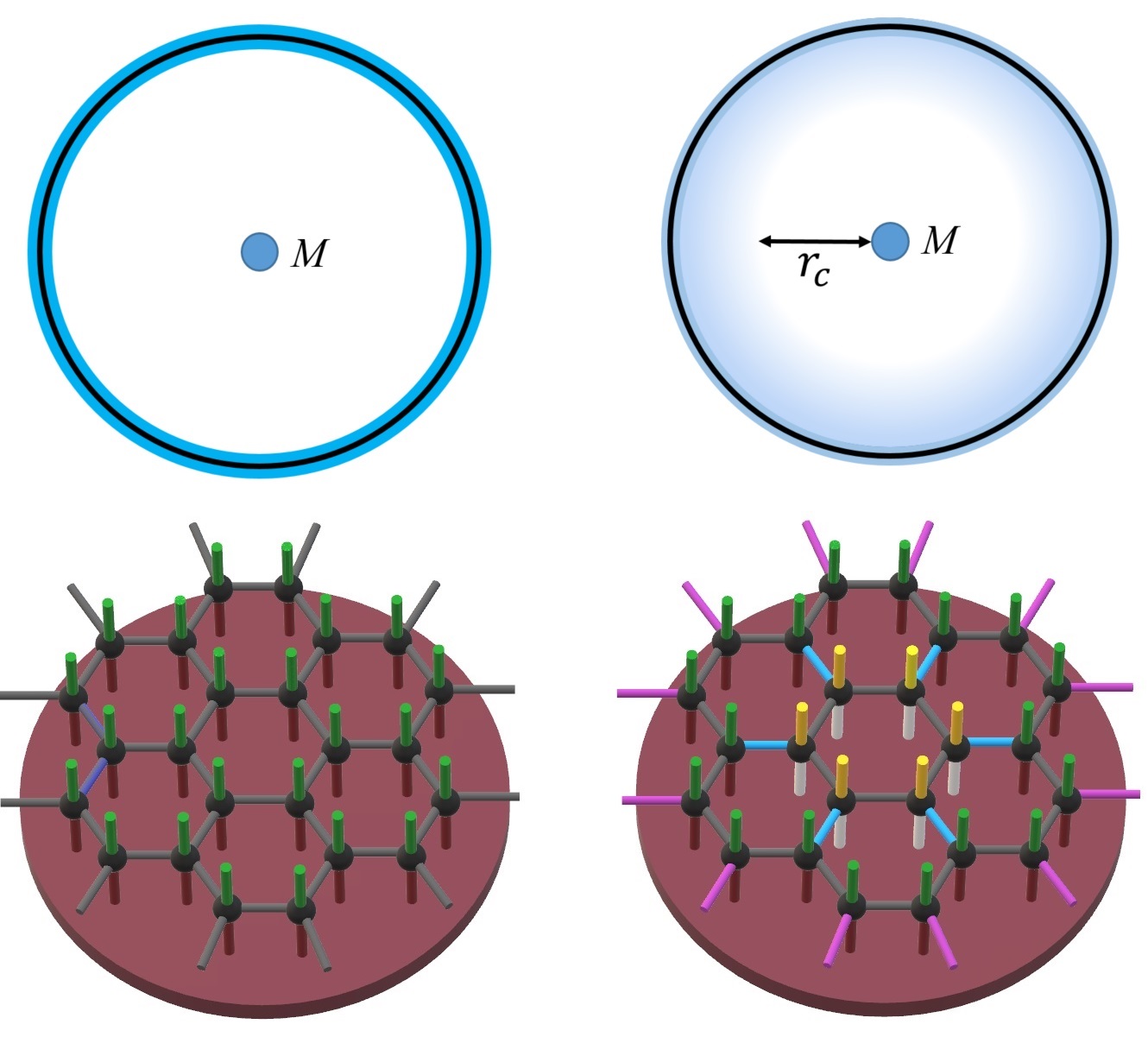}
\caption{(Top) Cartoon illustrating the encoding of the information (shaded blue) pertaining to the bulk mass $M$ in de Sitter space. For regions much smaller than $r_c$, the bulk information pertaining to $M$ is redundantly encoded in the boundary bits $N_\partial$ which obey an area-law. Conversely, for regions that are large compared to $r_c$, there are bulk degrees of freedom that scale extensively with the bulk and which are not encoded in the bits $N_\partial$. (Bottom) Cartoon of a portion of a tensor network representation for the entanglement structure of de Sitter space. The bulk legs correspond to the green indices. The entanglement structure that builds the emergent geometry is encoded in the short-range correlations described by the grey, internal legs of the network. The state is endowed with a constant density of long-range entanglement by the single, large red tensor. The bulk indices participate in the long-range entanglement via their connection to the large-red tensor via the dark red legs. (Bottom Right) adding matter locally removes some of the long-range entanglement. We imagine that the long-range indices in the center (white) have been deleted. In this case, the bulk state in the yellow legs can be approximately reconstructed from the state on the portion of network bounded by the cut which passes through the blue legs of the network. For larger portions of the network, such as for the cut which passes through the purple legs, the state in the bulk legs within this cut cannot be reconstructed from the legs passing through the cut, leading to a breakdown of holography for regions that are sufficiently large.}
\label{bigone}
\end{figure}

In \cite{verl2016}, the static patch of de Sitter space was considered, with the metric,
\begin{equation}
\label{dsstatic}
ds^2 = -f(r) dt^2 + \frac{1}{f(r)} dr^2 + d\Omega_{d-2}^2,
\end{equation}
with $f(r) = 1 - \frac{ r^2 }{ L^2 }$, where the cosmological horizon is at $r=L$. The Bekenstein-Hawking formula associates an entropy to de Sitter space which is determined by the area $A(L)$ of the cosmological horizon, as follows,
\begin{equation}
\label{dSent}
S_{DE} = \frac{ A(L) c^3 }{ 4 \hbar G },
\end{equation}
where, for reasons that will shortly be explained, the subscript ``DE" denotes the dark energy. In \cite{verl2016}, an interpretation for the entropy \eqref{dSent} is proposed wherein the total entropy of de Sitter is associated to the dark energy which is distributed throughout the volume of de Sitter, leading to constant entropy density that obeys a volume-law. Accordingly, if we consider a spherical region of size $r$, the entropy associated to the dark energy within this region is proportional to the volume $V(r)$ of the sphere, so that we have
\begin{equation}
\label{SV}
S_{DE}(r) = \frac{ V(r) }{ V_0 },
\end{equation}
where $V_0$ is the volume per unit of entropy of the dark energy. The condition that the entropy \eqref{SV} coincides with the de Sitter entropy \eqref{dSent} when $r=L$ then leads to the following formula for $V_0$,
\begin{equation}
\label{V0}
V_0 = \frac{ 4 G \hbar L}{ d-1 }.
\end{equation}
It is then straightforward to show that formula \eqref{dSent} can be re-written in the following way,
\begin{equation}
\label{SVA}
S_{DE}(r) = \frac{r}{L} \frac{ A(r) c^3}{ 4 G \hbar},
\end{equation}
which makes it clear to see that when $r=L$, the above formula reproduces the formula for the total de Sitter entropy \eqref{dSent}. In \cite{verl2016}, the emergence of dark gravity is traced to the interplay that results from the local removal of a portion of the dark energy degrees of freedom due to the addition of matter. To ascertain this effect, one can turn on a matter source for a point mass M at the origin of the static patch \eqref{dsstatic} by introducing the Newtonian potential with the following replacement,
\begin{equation}
\label{withpot}
f(r) \rightarrow 1 - \frac{ r^2 }{ L^2 } + 2 \phi(r),
\end{equation}
with,
\begin{equation}
\label{phi}
\phi(r) = \frac{- GM }{r }.
\end{equation}
It is shown that the negative sign of $\phi(r)$ leads to a reduction in the total de Sitter entropy when we add the mass $M$ to the origin of the static patch. This total reduction of the de Sitter entropy in fact corresponds to a local reduction of the entropy associated with the dark energy in the region surrounding the point mass. Accordingly one can consider the change in the growth of the area of a spherical region as a function of geodesic distance in the case with and without the matter to arrive at the following formula for reduction of the de Sitter entropy due to the addition of the mass $M$,
\begin{equation}
\label{SM}
S_M(r) = \frac{ 2 \pi M c}{\hbar} r.
\end{equation}
A key result of \cite{verl2016} is that, given a point mass M, there is a length-scale, depending on the mass and the curvature scale, where the de Sitter entropy \eqref{SVA} exactly equals the amount of entropy \eqref{SM} which is removed by the addition of a mass $M$. Using the formulas \eqref{SVA} and \eqref{SM} it is easy to calculate this scale to be the following,
\begin{equation}
\label{rcrit}
r_c(M) = \sqrt{\frac{ G M L}{c^2}} = \sqrt{\frac{ G M}{ a_0}},
\end{equation}
where the MOND acceleration scale $a_0 = c^2 / L$ \cite{milgrom} has been identified. Below this critical length-scale, all of the de Sitter entropy is removed by the mass $M$; in this case there are only matter degrees of freedom in the bulk which are encoded in the degrees of freedom at the boundary. This corresponds to what is called the Newtonian regime. Conversely, for regions that are larger than this scale, the mass $M$ does not remove all of the de Sitter entropy. In this sub-Newtonian or ``dark gravity" regime, there is therefore information associated to the dark energy in the bulk which is entangled with the bulk mass. This volume-law entanglement, which contains information about the bulk mass, then spoils the holographic encoding of the bulk mass. This scenario is depicted in figure \ref{bigone}. As we will see, this is a key feature of these proposals that this work seeks to emulate in the thermodynamic setup presented in section \ref{main}. 

In \cite{verl2016}, a remarkable feature is observed when we consider the following criteria, associated to the length-scale above which the volume-law associated with the dark energy overwhelms the matter entropy \eqref{SM}, that is where we have the following:
\begin{equation}
\label{entcrit}
S_M(r) < S_{DE}(r).
\end{equation}
Using equations \eqref{SVA} and \eqref{SM}, we observe that the above criteria are equivalent to the criteria \eqref{criteria}, as claimed in the introduction. 

Since the volume-law of the dark energy overwhelms the area-law scaling of the matter entropy in the dark gravity regime, then accordingly what emerges are not the known laws of gravity. Instead, in \cite{verl2016}, this regime is understood by considering the effective, elastic response of the dark energy medium to its removal, from local inclusion regions, by the addition of matter. It is shown that this leads to the fitting formula of MOND and also the baryonic Tully-Fisher relation \cite{milgrom, btf}. It is thereby argued that this framework provides an alternative explanation for the phenomena that are currently attributed to dark matter, which does not require dark matter to exist. This work will make contact with these two key results, but the effective elastic description described in \cite{verl2016} will not be required.

\section{Emergent Dark Gravity from a (Non)Holographic Screen}
\label{main}

This section describes how to obtain the emergence of a dark gravitational force, which scales like $\frac{1}{r}$ above a critical length-scale $r_c$, in a purely thermodynamic setting which is very similar to the setup considered in \cite{verl2010}. The modification presented here is inspired by the arguments in \cite{verl2016}, described in section \ref{review}, where the interplay between the entropy associated to the dark energy and the entropy which is removed by the addition of matter leads to a violation of the holographic encoding of bulk beyond the critical length-scale $r_c$ as in \eqref{rcrit}. Accordingly, the modification proposed here is the introduction of an arbitrary critical length-scale that controls the scale at which the holographic encoding of the bulk fails. 

\subsection{A (Non)Holographic Screen}
\label{nhs}

\begin{figure}
\centering
\includegraphics[width=\textwidth]
{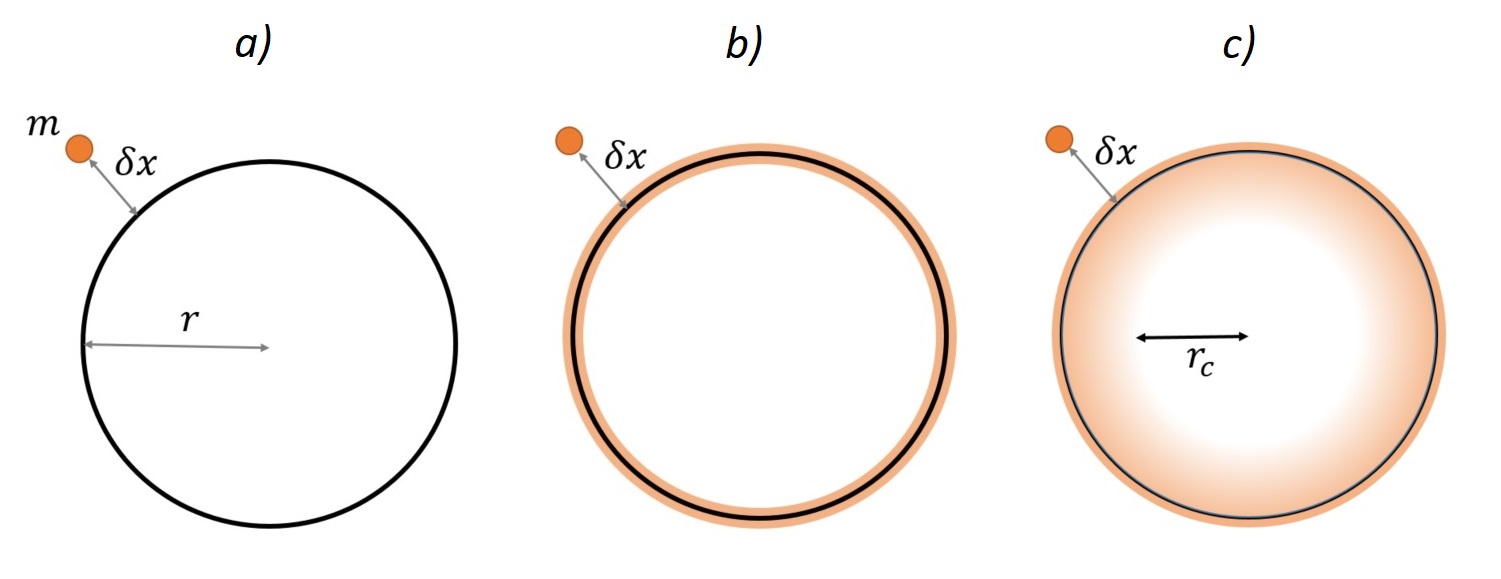}
\caption{ a) the test particle of mass $m$ is located in the emerged part of space, and is displaced from the spherical screen of radius $r$ by $\delta x \ll 1$. b) for regions $r \ll r_c$, the test particle contributes to a change in the entropy associated to the bits on the screen. c) for regions $r \sim r_c$, the test particle contributes to a significant change in the entropy associated to the bulk.}
\label{EDG}
\end{figure}

At first we will assume precisely the setup described in section 3.2 of \cite{verl2010}. Namely one is to imagine that there is a spherical region of space, known as a holographic screen, which separates the interior ``unemerged" part of space from the exterior ``emerged" part of space. A particular thermodynamic system is ascribed to this setup in which the ``unemerged" part of space emerges. The emergence of space is imagined to arise due to a series of coarse-graining steps that push the holographic screen into the ``unemerged" part of space, which leads to an overall reduction in the microscopic degrees of freedom associated with the holographic screen. In the emerged part of space, we imagine that there is a massive test particle of mass $m$ which is located at a small displacement $\delta x$ from the holographic screen. The setup is contrived so that a change in the test particle's position contributes a change to the entropy of the screen, which leads to an entropic force acting on the test particle. The setup so-far described is depicted in figure \ref{EDG} a). 

The role of the holographic screen, within the earlier work, is to encode the information within the unemerged space. To make a clear connection between this setup and the AdS/CFT literature, the unemerged part of space will be referred to, hereafter, as the bulk. The holographic screen then corresponds to the boundary of the bulk, but since this is an arbitrary boundary in space, and not a geometric boundary at spatial infinity as in AdS/CFT, the boundary of the bulk (unemerged part of space) will be referred to as a screen. In the present setup, holography will be explicitly violated, so the term ``holographic screen" will not be appropriate. The term non-holographic screen may then seem more appropriate, but for convenience this artificial boundary will simply be referred to as a screen.

In \cite{verl2010} it is assumed that the information pertaining to an emergent mass $M$ in the bulk is encoded holographically in a quantity of bits $N_\partial$ which is proportional to the area of the screen, in the following way,
\begin{equation}
\label{nbound}
N_\partial = \frac{ A(r) c^3}{ 4 G \hbar}.
\end{equation}
The main modification of \cite{verl2010} presented in this paper is to assert that the holographic principle is violated in a way that depends on a critical length-scale $r_c$. We suppose that, in addition to the bits $N_\partial$, there are a set of bits $N_\Sigma$ that one also needs to have in order to be able to completely reconstruct the emergent mass $M$ in the bulk. We assert that the bits $N_\Sigma$ pertaining to $M$ are not encoded holographically at the screen. Motivated by \cite{verl2016} we assert that $N_\Sigma$ should scale extensively with the \emph{volume} of the bulk. We then associate the critical scale $r_c$ with the ratio of the numbers of bulk and screen bits as follows,
\begin{equation}
\label{ratio}
\frac{N_\Sigma}{N_\partial} = \frac{r}{r_c},
\end{equation}
which leads to the following form for $N_\Sigma$,
\begin{equation}
\label{BV}
N_\Sigma = \frac{r}{r_c} \frac{ A(r) c^3}{ 4 G \hbar}.
\end{equation}
Based on the above observations we conclude that the role of the critical scale $r_c$ is to determine when the holographic, area-law scaling of the information associated to $M$ in the bulk is overwhelmed by the volume-law contribution of the bulk bits $N_\Sigma$ that are not encoded holographically. Thus the critical scale $r_c$ that we have described plays a very similar role to the critical scale $r_c$ as in \eqref{rcrit} identified in \cite{verl2016}. In our setup, one could suppose that the scale $r_c$ that we are considering, should depend on the amount of information associated to the mass $M$ and the curvature scale $L$. This is indeed the case for the critical scale \eqref{rcrit}. For now, this dependence will not be explicitly assumed.

If one imagines inserting a mass into the bulk, we can call the number of bits that characterise this matter $\tilde{N}_\Sigma$. In general this is different from $N_\Sigma$, because when the holographic principle holds, we anticipate that the former are encoded redundantly at a screen, of size $r$, as a quantum secret-sharing scheme among the bits $N_\partial(r)$ \cite{harlow14, harlow15, harlow16}. In contrast, the bits $N_\Sigma$ pertain to information about the bulk mass that is not encoded holographically, which do not therefore form a subset of the bits $N_\partial$. The latter can be compared to the formula \eqref{SVA}, so that the bits $N_\Sigma$ here play the same role as the dark energy in \cite{verl2016}. 

\subsection{Thermodynamic Setup}

Now we describe how to associate a thermodynamic system with the setup described in \ref{nhs}. As per \cite{verl2010}, we begin with the assumption that displacing the test particle which is outside the screen, but which is close to it, leads to a change to the entropy of the screen and the bulk. The change in the entropy of the total system will take the following form,
\begin{equation}
\label{composite}
\delta S = \delta S_\partial + \delta S_\Sigma ,
\end{equation}
where $\delta S_\partial$ coincides with $\delta S$ defined in section 3.2 of \cite{verl2010}, which we identify as the change in the entropy of the screen. The quantity $\delta S_\Sigma$ is similarly defined as the contribution to the entropy of the bulk, which, as we have explained, is not encoded holographically. 

A change to the entropy as per \eqref{composite} will lead to a change to the energy of the system $\delta E = \delta E_\partial + \delta E_\Sigma$, where we have identified the contributions $\delta E_\partial$ and $\delta E_\Sigma$ which correspond to changes to the energy associated with the screen and bulk degrees of freedom respectively. Following \cite{verl2010}, we assume that when the displacement $\delta x$ of the test particle is small, the corresponding changes to the energies of the bulk and screen systems should be evenly divided over their respective degrees of freedom, so that the equipartition theorem holds for each system.

\begin{equation}
\label{equipartitions}
\delta E_\partial = \frac{1}{2} N_\partial k_B T_\partial \hspace{1cm} \delta E_\Sigma = \frac{1}{2} N_\Sigma k_B T_\Sigma ,
\end{equation}
where $T_\partial$ and $T_\Sigma$ are the temperatures associated to the bulk and boundary systems. To proceed we further make the assumption that the bulk and boundary systems are in thermal equilibrium. In general this is something that we may expect to hold for situations where the emergent bulk is static. For now we will assume that this holds, so that we can identify $T_\partial = T_\Sigma = T$, where $T$ is the equilibrium temperature associated to the combined ensemble of the bulk and the screen. 

If the first-law of thermodynamics holds for this system, then for near-equilibrium configurations the following relation must hold for the bulk and boundary subsystems,
\begin{equation}
\label{1stlaw}
\delta E_\partial =T \delta S_\partial \hspace{1cm} \delta E_\Sigma = T \delta S_\Sigma .
\end{equation}
Using equations \eqref{equipartitions} and \eqref{1stlaw}, we find that the changes to the entropy of the bulk and the screen obey the following relation,
\begin{equation}
\label{heidi}
\delta S_\Sigma = \frac{N_\Sigma}{N_\partial} \delta S_\partial = \frac{r}{r_c} \delta S_\partial .
\end{equation}
The above relation has the intuitive interpretation that for scales $r \ll r_c$, we have $\delta S_\Sigma \sim 0$, and the significant contribution to the entropy of the system comes from the contribution to the screen, as depicted in figure \ref{EDG} b). Conversely, at $r=r_c$, the test particle contributes the same amount of information to the bulk and boundary subsystems at $r=r_c$. Hence $r_c$ naturally described a scale where the contribution of a test particle to the bulk first becomes comparable to its corresponding contribution to the screen. For scales $r>r_c$, the corresponding contribution to the bulk overwhelms the contribution to the screen, as depicted in figure \ref{EDG} c). Putting together equations \eqref{heidi} and \eqref{equipartitions} gives the following relation between the changes to the energies of the bulk and boundary subsystems,
\begin{equation}
\label{heidi2}
\delta E_\Sigma = \frac{N_\Sigma}{N_\partial} \delta E_\partial = \frac{r}{r_c} \delta E_\partial .
\end{equation}
To obtain an emergent entropic force associated to this system, we express the change in energy given by the first law \eqref{1stlaw} in terms of the work done by the force $F$ that displaces the test particle a distance $\delta x$ \emph{towards} the screen. In this way the entropic force must take the following form in terms of the change to the entropy of the system.
\begin{equation}
\label{ef}
F = T \Bigg( \frac{\partial S_\Sigma}{\partial x}+\frac{\partial S_\partial}{\partial x} \Bigg) = T \frac{\partial S_\partial}{\partial x}\Bigg( 1 + \frac{r}{r_c} \Bigg),
\end{equation}
where equation \eqref{heidi} was used in the last step. The term $\frac{r}{r_c}$ includes a modification to the setup described in \cite{verl2010} which is associated with the breakdown of holography that we have described. In particular, when $r \ll r_c$ we can effectively set this term to zero, and in this case we would obtain the same result as \cite{verl2010}, where the holographic principle is assumed to hold exactly.

\subsection{Emergent Dark Gravity}
\label{mainmain}

In \cite{verl2010}, the form for the change $\delta S_\partial$ to the entropy of the screen can be motivated as follows. Following Bekenstein's derivation of the black hole entropy, one can propose that when the test particle is one Compton-wavelength from the screen, it should contribute a single bit of information to it \cite{bek}. This contribution is assumed to be linear in $\delta x$, at least approximately when $\delta x$ is sufficiently small. This leads to the following form for the change of the boundary entropy,
\begin{equation}
\label{ds1}
\delta S_\partial = \frac{\pi k_B}{2} \frac{ m c }{\hbar } \delta x .
\end{equation}
To relate the quantities in this thermodynamic system to an emergent mass we suppose that the following relations hold,
\begin{equation}
\label{firstlaw}
\delta E_\partial =M_B c^2 \hspace{1cm} \delta E_\Sigma = M_D c^2 .
\end{equation}
For now $M_B$ and $M_D$ are just arbitrary constants with the units of mass. The relation \eqref{heidi2} implies that these quantities obey the relation,
\begin{equation}
\label{BTF}
\frac{M_D}{M_B} = \frac{r}{r_c} .
\end{equation}
We can now use either of the equations \eqref{equipartitions} to determine the equilibrium temperature $T$ which is found to be the following,
\begin{equation}
\label{combtemp}
T = \frac{ 2 M_B G \hbar}{c k_B \pi r^2} .
\end{equation}
Now we have all of the pieces to obtain the form for the entropic force that emerges from this thermodynamic system. In the case where $r \ll r_c$, the bulk contribution to the entropy change can be neglected. In this limit we can neglect our modification to the setup described in \cite{verl2010} and we accordingly obtain the same result contained therein, which is the emergence of Newton's gravitational force-law,
\begin{equation}
\label{Enewton}
F = \frac{ G m M_B }{ r^2 } .
\end{equation} 
Equation \eqref{Enewton} justifies the interpretation of $M_B$ as describing an emergent mass in the bulk. For $r \sim r_c$, the bulk contribution to the change in entropy of the screen becomes non-trivial, and using the formula \eqref{combtemp} for the equilibrium temperature, together with the formulae \eqref{heidi} and \eqref{ds1} for the changes in the bulk and boundary entropy (respectively) and plugging these results into \eqref{ef} for the entropic force, the entropic force takes the form,
\begin{equation}
\label{darkgravity}
F = \frac{ G m M_B }{ r^2 } + \frac{G M_B m}{ r \cdot r_c} .
\end{equation}
This force has the correct $\frac{1}{r}$ scaling required to describe flat galactic rotation curves, where $M_B$ is the baryonic point mass located at the origin of the bulk. It is worth emphasising that this result essentially derived from the violation of holography as per equation \eqref{ratio}, which eventually led to equation \eqref{heidi}. In view of the derivation of Newtonian gravity in section 3.2 of \cite{verl2010}, the additional factor of $r$ contained in \eqref{heidi}, in this work, is what led to a force that scales like $r^{-1}$.

Presently, the critical scale appearing in \eqref{darkgravity} is arbitrary. On the other-hand, the critical length-scale \eqref{rcrit} plays essentially the same role in \cite{verl2016}. If we identify our critical scale with the scale \eqref{rcrit}, we obtain the following emergent force,
\begin{equation}
\label{DEG}
F= \frac{G M_B m}{r^2} + \sqrt{M_B G a_0} \frac{m}{r},
\end{equation}
which has precisely the form of Newtonian gravity with an additional dark gravity force which is observed at galactic scales above $r_c$ as per equation \eqref{rcrit} \cite{milgrom}. 
 By inserting the relation \eqref{BTF} into \eqref{darkgravity} we see that the entropic force can be written in the following form,
\begin{equation}
\label{darkgravity2}
F = \frac{ G m }{ r^2 }  \big( M_B + M_D(r) \big) .
\end{equation}
This result serves to clarify the role of the quantity $M_D$, which describes the change in the bulk energy according to \eqref{firstlaw}. In view of \eqref{darkgravity2}, we conclude that the failure of holography for scales $r \sim r_c$ leads to a contribution to the bulk energy (of the test particle to the screen) that manifests as an apparent, additional mass $M_D$ in the bulk. Notice that the scale $r_c$ drops out of equation \eqref{darkgravity2}, but to ascertain the relation between the apparent mass $M_D$  and the mass $M_B$, we need to use the relation \eqref{heidi2}, which does implicate $r_c$. This setup can therefore mimic the dark matter hypothesis, independently of the choice of $r_c$. We see that the identification between the mass $M_B$ and the apparent mass $M_D$ given by equation \eqref{heidi2} has a similar role to the baryonic Tully-Fisher relation, which relates the baryonic mass distribution to the apparent distribution of dark matter \cite{btf}. 

If we consider the emergence of a spherically-symmetric mass distribution in the bulk which is entirely contained in a screen of size $r$. That is, suppose that there is an emergent mass profile $M_B(r)$ in the bulk, contained inside a screen of size $r$ and which corresponds to the change in energy $\delta E_\partial$ of the screen. Then if we then take the formula \eqref{BTF}, with the critical scale $r_c$ identified by \eqref{rcrit}, we recover a relation between the emergent mass profile $M(r)$ and the apparent mass profile $M_D(r)$, which, up to a numerical factor, has been shown to be equivalent to the baryonic Tully-Fisher relation \cite{btf},
\begin{equation}
\label{BTFfinal}
M_D(r)^2 = \frac{a_0}{2 \pi G} M_B(r),
\end{equation}
which was shown to determine the apparent dark matter profile $M_D(r)$, given a profile $M_B(r)$ of (observed) baryonic matter in galaxies. Using this framework, we can therefore make contact with the main results obtained in \cite{verl2016}. 

\subsection{Review of Key Assumptions}
The key features of our main result, presented in \ref{mainmain}, follow from the application of the following assumptions, whose role should be clarified,
\begin{enumerate}
\item{The bulk is not encoded holographically at the screen, and the non-holographic bulk degrees of freedom obey a volume-law, such that the failure to encode the bulk holographically is controlled by a length-scale $r_c$.}
\item{Changing the displacement of the test particle changes the entropy associated with the bulk and boundary subsystems.}
\item{The change in energy of the bulk and boundary subsystems obeys the equipartition theorem.}
\item{The bulk and boundary systems are in thermal equilibrium.}
\item{The first law of thermodynamics holds.}
\end{enumerate}
Assumption 1 is our main assumption in this work, and it is the main modification to the setup described in \cite{verl2010}. This is the modification that makes a connection to the new theory \cite{verl2016}, where a violation of the holographic principle occurs for length-scales which are sufficiently large so that the inequality \eqref{entcrit} holds. Assumption 2 is also made in \cite{verl2010}, and this is the key feature that is contrived to produce an entropic force acting on the test particle. Assumption 3 is also made in \cite{verl2010} in relation to what we have called the boundary system. That we should imagine this to hold for the bulk system seems like a natural extension, since when the particle is sufficiently close to the screen, we can imagine that the change in the energy of the bulk is evenly divided over the bulk bits. As in \cite{verl2010}, we may not expect this to hold for displacements that are large compared to the Compton wavelength of the test particle, and we refer the reader to the former work for a justification of this. Assumption 4 is new here and it seems like a natural assumption to make when the emergent bulk and matter distribution is static, as was the case for the static patch of de Sitter that was considered in \cite{verl2016}. In more general situations, where there are non-trivial dynamics, we may not expect assumption 4 to hold exactly. Assumption 5 was assumed in \cite{verl2010}; the first law should evidently hold for any equilibrium thermodynamic system, such as the one that has been described here. The role of the first law is, as in the earlier work, that it allows one to relate the an entropic force, to the entropy gradient and the equilibrium temperature as per \eqref{ef}.

In many cases, it may appear that the fundamental constants have been introduced in an ad-hoc fashion. This objection could also be raised about the work \cite{verl2010}. Here, as in that work, the role of these contants is essentially to give quantities with the correct dimensions. Nevertheless, we see that $\hbar$ drops out in the calculations of the dark gravity force \eqref{DEG}, as it must in order to match with the idea that we are considering a Newtonian limit (that does not albeit lead to Newtonian gravity). So, as with the previous work, the constant $\hbar$ remains arbitrary in this work.

\section{Comparison with Previous Work}
\label{relation}

The setup described in section \ref{main} is strongly based on the setup described in \cite{verl2010}. However, a key modification is included which is based on a central observation of the recent proposals \cite{verl2016}, that the emergence of dark gravity is due to the breakdown of the holographic principle above a certain length-scale \eqref{rcrit}, where the inequality \eqref{entcrit} holds, as described in section \ref{review}. Indeed, one of the aims of this work is to make a clear connection between the relatively simpler framework described in \cite{verl2010} and the recent work \cite{verl2016}, by extracting this key feature and implementing it as we have described in section \ref{main}. The main results \eqref{DEG} and \eqref{BTFfinal} that we have derived for the emergent dark gravity force and the baryonic Tully-Fisher relation (respectively), rely on the identification for our critical scale as per \eqref{rcrit}. We cannot motivate the use of this scale based on the efficacy of the empirical observations because in the present work, the role of the critical scale is to control the breakdown of holography, which does not immediately follow from the empirically observed behaviour above this scale. This effect is indeed currently widely attributed to the dark matter hypothesis, which is an entirely different proposal. Thus, the motivation for us to consider the scale \eqref{rcrit}, in the present work, is tied to the proposals of \cite{verl2016} which initially make contact with the observed criteria. In that work, this length-scale is motivated by according a particular interpretation to the de Sitter entropy, which it not considered here; it is instead shown that the ensuing violation of holography, with a volume-law scaling of the emergent matter degrees of freedom, essentially leads to the key features that match the observed galaxy-scale phenomena. To this end the effective elastic description that was described in \cite{verl2016} was not required.

There have been several recent papers with related goals, which have considered how to derive the MOND fitting formula in terms of an entropic force which emerges from a thermodynamic setup involving holographic screens. These works include \cite{Zhang}, where an argument analogous to \cite{verl2010} is considered, but with the introduction of a modified inertia relation which accordingly produces a modified gravity force. In this work we do not assume that such a modified inertial relation holds. Rather, as we have said, the present setup is inspired by the observation of Verlinde in \cite{verl2016} that the emergence of a dark gravity force is attributed to a breakdown of holography at a certain scale, which has been implemented explicitly in setion \ref{main} of the present paper. There are also a pair of more recent papers \cite{Abreu1,Abreu2} which have considered how to derive the MOND relation, again inspired by the holographic screen arguments \cite{verl2010}, but the authors have made use of the Tsallis entropy to derive a modified gravity relation. Again, this differs significantly from the present approach, where Tsallis statistic have not been used; the key results presented here derive from a breakdown of holography which is not assumed in the previous papers \cite{Abreu1,Abreu2}.

\section{Discussion \& Outlook}

In this work, a clear connection has been established between the earlier work \cite{verl2010} and the recent proposals \cite{verl2016}, by implementing a breakdown of holography which is controlled by a critical length-scale via a straightforward modification to the earlier framework. Furthermore, when the value for the critical scale \eqref{rcrit}, identified in the recent proposals, is adopted, the exact form of the dark gravity force, in addition to (up to a numerical factor) the baryonic Tully-Fisher relation, which are observed to produce flattening galactic rotation curves above precisely this scale, are obtained. This work then essentially clarifies the key role played by the breakdown of holography for sufficiently large length-scales (given an emergent mass $M$) in E. Verlinde's new hypothesis, whilst providing a mathematically simple framework with which one could explore this, and related ideas. 

In figure \ref{bigone}, a tensor network has been depicted which provides an analogy for the entanglement structure of an emergent de Sitter geometry as described in \cite{verl2016}. The particular network represented here does not accurately depict the tensor network one might use to describe an emergent de Sitter geometry, however one could in principle obtain this via a discretisation of a constant time slice \cite{B16, ev}, sewn with tensors that, as well as bulk and internal indices, each carry a small additional index which is contracted with the tensor that thereby endows the state with a constant density of long-range entanglement, whose role can therefore mimic the dark energy as per \cite{verl2016}. Furthermore, this work offers the interpretation that the addition of a mass $M$ can be framed, in these terms, as a deletion of a portion of the long-range legs attached to a closed subregion of the network. This would be consistent with the idea that mass is associated with relative entropy in emergent gravity\cite{verl2016}. This question is left open for future work.

 A particularly interesting open question for this work relates to the possibility of a finding covariant formulation, which could make contact with the recent work \cite{sabine, wang}. Most of the components needed for this have been identified in the present work, but the formulation of a covariant model is left for future work. Another interesting and partially-related question  concerns a possible relation between the framework presented here and the effective elastic description offered in \cite{verl2016} to describe the dark gravity regime.

Despite the many successes of the dark matter paradigm, we have presently yet to observe the dark matter particle, and Verlinde's new theory offers the exciting possibility that understanding this regime may require a radical revision of our widely-held belief in the efficacy of GR and EFT on cosmological scales. In this work we hope to have provided a framework that will help to further our understanding of this new proposal and its underlying microscopic description.


\section{Acknowledgements}
The author would like to thank Henry Maxfield and Vaios Ziogas for helpful discussion and comments.

\bibliographystyle{JHEP}
\bibliography{EDGfinal}

\end{document}